In loving memory of a wonderful grandma whose courageous battle with cancer motivated me to write this paper

# Efficient Brain Tumor Segmentation Using a Dual-Decoder 3D U-Net with Attention Gates (DDUNet)

Mohammad Mahdi Danesh Pajouh[1]

Independent Research

## Abstract

Cancer remains one of the leading causes of mortality worldwide, and among its many forms, brain tumors are particularly notorious due to their aggressive nature and the critical challenges involved in early diagnosis. Recent advances in artificial intelligence have shown great promise in assisting medical professionals with precise tumor segmentation, a key step in timely diagnosis and treatment planning. However, many state-of-the-art segmentation methods require extensive computational resources and prolonged training times, limiting their practical application in resource-constrained settings.

In this work, we present a novel dual-decoder U-Net architecture enhanced with attention-gated skip connections, designed specifically for brain tumor segmentation from MRI scans. Our approach balances efficiency and accuracy by achieving competitive segmentation performance while significantly reducing training demands. Evaluated on the BraTS 2020 dataset, the proposed model achieved Dice scores of 85.06% for Whole Tumor (WT), 80.61% for Tumor Core (TC), and 71.26% for Enhancing Tumor (ET) in only 50 epochs, surpassing several commonly used U-Net variants. Our model demonstrates that high-quality brain tumor segmentation is attainable even under limited computational resources, thereby offering a viable solution for researchers and clinicians operating with modest hardware. This resource-efficient model has the potential to improve early detection and diagnosis of brain tumors, ultimately contributing to better patient outcomes.

---

[1] Mohammadmahdi.danesh@ucalgary.ca

# 1. Introduction

## 1.1 Clinical Introduction

Brain tumors are abnormal growths of cells within the brain or its surrounding structures. They can occur in various regions, including the brain tissue itself, its protective lining, the brainstem, and even the nasal cavity or sinuses. Due to the brain's complex anatomy and the critical functions of its regions, tumors can lead to significant neurological impairment by compressing or invading healthy tissue, disrupting fluid circulation, or increasing intracranial pressure.

In the United States, brain and nervous system tumors affect approximately 30 out of every 100,000 adults. These tumors are classified as either benign (non-cancerous) or malignant (cancerous). While benign tumors generally grow slowly and do not spread, they can still be life-threatening depending on their location and size. Malignant tumors, on the other hand, tend to grow rapidly and infiltrate nearby healthy brain tissue, posing a serious risk to life and function.

Brain tumors may originate within the brain (primary tumors) or result from metastasis of cancers from other parts of the body (secondary or metastatic tumors), the latter being significantly more common. Cancers such as lung, breast, melanoma, kidney, and colon cancers frequently metastasize to the brain. The diversity and severity of brain tumors underscore the importance of early and accurate diagnosis to guide treatment and improve outcomes [1].

## 1.2 Importance of Early Detection

Early detection of brain tumors is critical for improving patient outcomes. It significantly enhances the likelihood of survival, particularly among vulnerable populations such as children and young adults, where brain tumors are a leading cause of cancer-related morbidity. Prompt recognition of symptoms can prevent tumor progression and enable timely medical intervention.

Detecting tumors in their early stages can preserve cognitive function by preventing irreversible brain damage. It also opens the possibility for less aggressive treatment options, such as surgery alone, and facilitates the use of targeted therapies with fewer side effects. Additionally, early-stage treatment is often more cost-effective, reducing the financial burden on patients and healthcare systems.

Overall, early diagnosis not only increases survival rates but also improves quality of life by minimizing neurological deficits and enabling more favorable treatment outcomes [2].

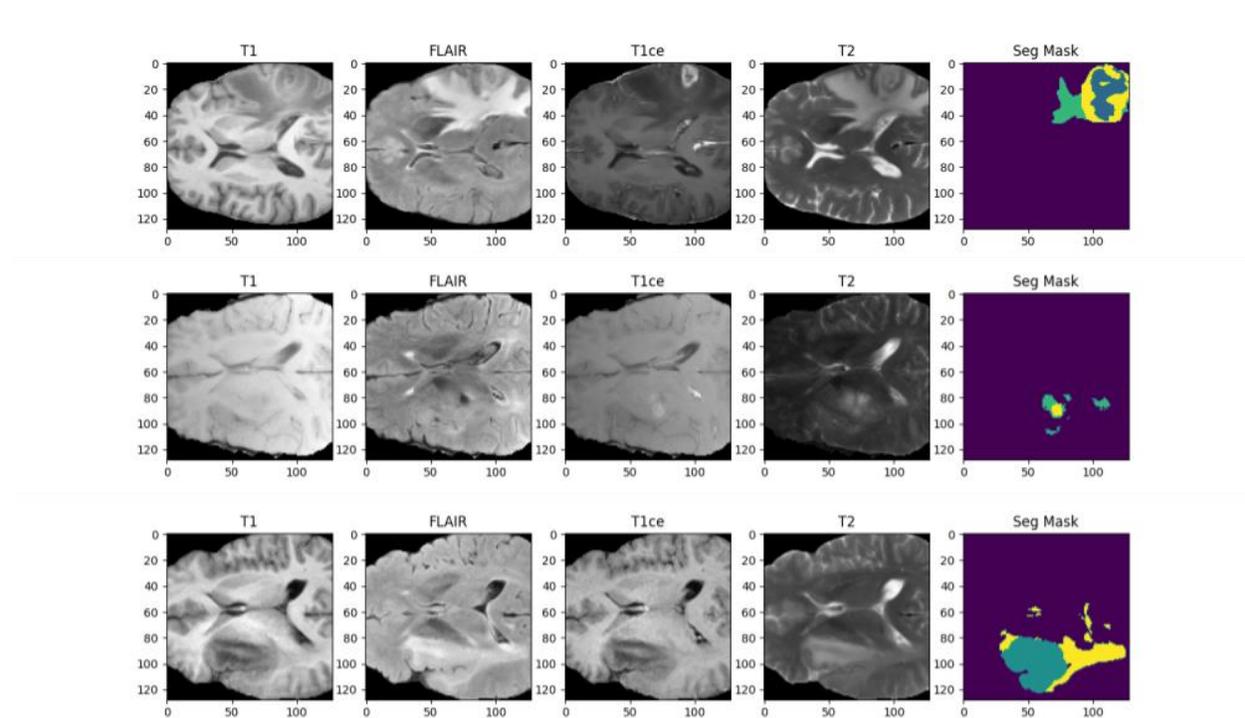

*Figure 1 Visualization of multi-modal MRI scans from the BraTS 2020 dataset. Each row displays axial slices of a single patient across four MRI modalities—T1-weighted (T1), Fluid-Attenuated Inversion Recovery (FLAIR), post-contrast T1-weighted (T1ce), and T2-weighted (T2)—followed by the corresponding segmentation mask. These complementary modalities provide diverse structural and contrast information critical for accurate brain tumor segmentation.*

### 1.3 Magnetic Resonance Imaging

Magnetic Resonance Imaging (MRI) is a non-invasive imaging technology that produces three dimensional detailed anatomical images. It is often used for disease detection, diagnosis, and treatment monitoring. It is based on sophisticated technology that excites and detects the change in the direction of the rotational axis of protons found in the water that makes up living tissues [3].

Unlike X-ray based methods, MRI does not involve ionizing radiation and excels in producing images with excellent soft-tissue contrast. This capability is particularly valuable for visualizing the brain's intricate anatomy. In clinical brain studies, multiple MRI modalities are typically acquired to emphasize different tissue characteristics.

In brain tumor imaging, common MRI sequences include T1-weighted (T1), T1-weighted with contrast enhancement (T1ce), T2-weighted (T2), and Fluid-Attenuated Inversion Recovery (FLAIR) shown in figure 1. T1-weighted images are used for capturing fine anatomical details, while T1ce images—obtained after administering a gadolinium-based contrast agent—highlight regions of increased blood flow or vascularity, such as tumors, abscesses or areas of inflammation tissue. T2-weighted images are particularly sensitive to fluid, making them useful for identifying edema, which is the accumulation of excess fluid in brain tissue surrounding a tumor. FLAIR sequences suppress the signal from cerebrospinal fluid, enabling clearer visualization of lesions adjacent to fluid spaces. These complementary imaging modalities collectively provide a robust basis for characterizing brain tumors.

MRI scans produce three-dimensional (3D) volumetric data composed of multiple 2D slices. To analyze this data, it is common to view the images in three anatomical planes depicted in figure 2:

- Axial (Horizontal) View: Slices are taken parallel to the ground, from the top of the head down to the base. This is one of the most commonly used views for brain imaging, allowing visualization of structures from superior to inferior.
- Coronal (Frontal) View: Slices are oriented vertically, dividing the brain into front and back portions. This view is helpful for examining structures from the anterior to posterior side.
- Sagittal View: Slices run vertically from left to right, dividing the brain into left and right halves. This view is particularly useful for observing midline structures and asymmetries.

Together, these views provide comprehensive spatial context necessary for accurate interpretation and segmentation of brain tumors.

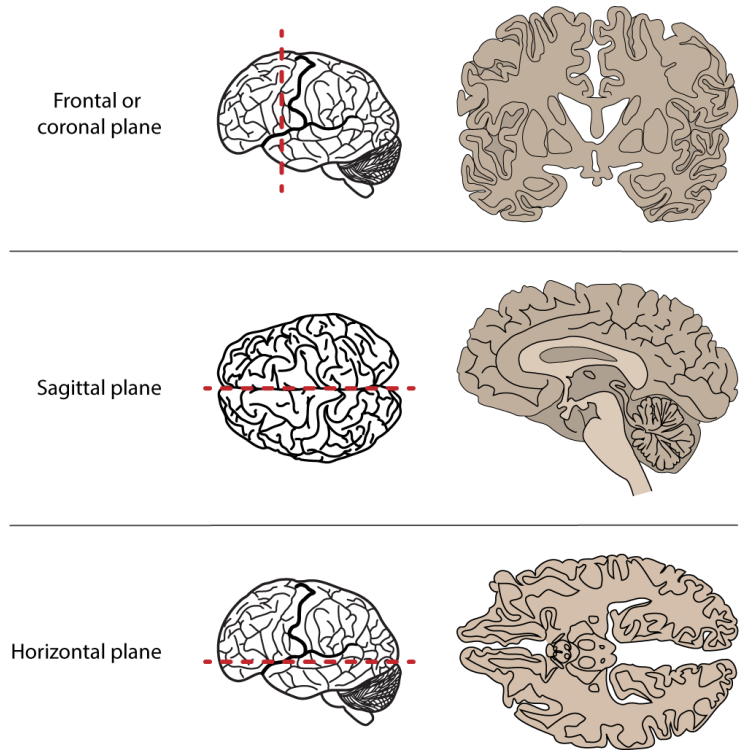

*Figure 2 Visualization of the three standard anatomical planes used in MRI imaging. (Top) The frontal (or coronal) plane slices the brain vertically from side to side, dividing it into anterior and posterior (front and back) sections. (Middle) The sagittal plane divides the brain vertically from front to back, separating it into left and right halves. (Bottom) The horizontal (or axial) plane cuts horizontally through the brain, yielding top-down slices from superior to inferior regions. These views provide complementary perspectives for comprehensive evaluation of 3D brain structures in clinical and research applications. The image is adapted from "Foundations of Neuroscience" by Casey Henley (2021), licensed under CC BY-NC-SA 4.0.*

## 1.4 Dataset

The BraTS 2020 [4,5,6] dataset is a well-established benchmark for brain tumor segmentation research. It consists of multi-modal MRI scans along with detailed manual annotations that delineate critical tumor subregions. In these annotations, distinct tissue characteristics are highlighted—such as regions of active tumor where the tissue is enhancing, areas of necrosis (cell death within the tumor), and zones of peritumoral edema. Edema indicates the spread of fluid around the tumor and can significantly affect brain function, while necrosis reflects aggressive tumor behavior and cell death. The dataset labels are generally defined as follows: label 0 corresponds to the background; label 1 represents necrotic and non-enhancing tumor core; label 2 indicates peritumoral edema; and label 3 denotes enhancing tumor. Researchers often combine these labels to define clinically relevant regions, such as the whole tumor (the union of labels 1, 2, and 3), tumor core (labels 1 and 3), and the enhancing tumor (label 3). By providing these comprehensive annotations, the BraTS dataset enables consistent evaluation of segmentation algorithms and supports the development of models that can accurately differentiate between various tumor components.

### 1.5 Proposed model: Motivation and Rationale

After extensive testing of different architectures, we introduce Attention Dual-Decoder 3D-Unet or Attention DDUNet for short for the task of brain tumor segmentation. The 3D architecture allows the model to learn 3D dependencies in volumetric MRI images. Having two separate decoders, relies on stochastic nature of training to learn diverse features that lead to better generalization. While the two separate attention mechanisms provide decoders with salient features from encoder to achieve better segmentation results. The models are evaluated based on dice score, sensitivity and specificity.

The motivation behind proposing this method is threefold, reflecting both practical and scientific goals.

**1. Enhancing a widely-used baseline (U-Net):**

U-Net is among the most influential and widely adopted architectures in biomedical image segmentation due to its encoder–decoder structure and skip connections, which help preserve spatial information. Despite the introduction of more sophisticated models in recent years, U-Net and its variants remain a strong baseline for medical segmentation tasks, including brain tumor segmentation. This work introduces a dual-decoder U-Net-based architecture that significantly outperforms the standard U-Net and its variants, particularly when constrained to limited training epochs. By integrating architectural adjustments such as separate attention gates for each decoder, the proposed model is able to extract richer and more diverse features without substantial computational overhead.

The proposed architecture is the result of an extensive empirical investigation. Over one hundred different combinations of network structures, hyperparameters, and training strategies were systematically explored and evaluated. This exhaustive experimentation aimed to identify a model that not only offers strong segmentation performance but also maintains low computational requirements. Among the tested configurations, a dual-decoder U-Net enhanced with attention gates emerged as the most effective, consistently outperforming its counterparts on the BraTS 2020 dataset.

**2. Accessibility and hardware efficiency:**

Another key motivation is to develop a model that remains highly accessible to researchers, clinicians, and developers working with limited hardware. Many high-performing models in literature rely on GPUs with large memory capacity and extended training times, making them inaccessible to a large segment of potential users. This model, in contrast, has been developed and evaluated under modest hardware constraints, such as a standard home PC. Despite these limitations, it achieves strong performance, demonstrating that high-quality segmentation is possible without requiring high-end infrastructure. This accessibility can empower researchers to fine-tune or deploy the model on their own datasets with minimal effort and resources.

**3. Realistic evaluation and clinically relevant metrics:**

Finally, this work emphasizes practical and meaningful evaluation metrics. While some prior works have reported high overall accuracy—sometimes exceeding 98%—such figures can be misleading in segmentation tasks due to class imbalance. In particular, metrics like accuracy can be artificially inflated through micro-averaging, where the abundance of background pixels dominates the score. Instead, this study prioritizes macro-averaging and clinically relevant metrics such as the Dice Similarity Coefficient (DSC), which better reflect segmentation performance on smaller but critical regions. This choice ensures that the reported results are aligned with real-world performance expectations in clinical applications.

In summary, this study introduces a dual-decoder U-Net-based model that improves upon common segmentation baselines like U-Net, particularly in low-resource settings. The method is designed for accessibility, allowing researchers to train and deploy it on modest hardware. It emphasizes meaningful clinical evaluation using Dice score, ensuring realistic and practical performance metrics.

## 2. Related Works

Deep learning has significantly advanced brain tumor segmentation, with U-Net and its variants playing a central role. These architectures have evolved to address the complex challenges inherent in medical imaging, such as varying tumor shapes, sizes, and locations.

Introduced by Ronneberger et al., U-Net employs an encoder-decoder structure with skip connections, enabling precise localization in biomedical image segmentation tasks [7]. Its design allows for the combination of contextual information from the encoder with high-resolution features from the decoder, facilitating accurate segmentation even with limited training data.

To handle volumetric data inherent in medical imaging, Çiçek et al. extended U-Net to 3D, facilitating effective brain tumor segmentation in MRI scans [8]. This adaptation processes 3D volumes directly, preserving spatial context across all dimensions and improving segmentation accuracy for complex structures like tumors.

Attention U-Net incorporates attention gates into the U-Net framework, allowing the model to focus on relevant features and suppress irrelevant ones. In the Attention U-Net architecture, attention gates are placed at the skip connections between the encoder and decoder paths. These gates use additive soft attention mechanisms to weigh the importance of features from the encoder before merging them with decoder features. This process allows the network to suppress irrelevant background information and highlight salient features pertinent to the segmentation task [9].

The E1D3 U-Net architecture extends the standard 3D U-Net by incorporating one encoder and three decoders, each dedicated to segmenting a specific tumor sub-region: whole tumor, tumor core, and enhancing tumor. This design allows for specialized feature learning in each decoder, improving segmentation performance across different tumor components [10].

nnU-Net (short for "no-new-Net") is an open-source, self-configuring deep learning framework specifically designed for biomedical image segmentation tasks. Unlike traditional segmentation models that require manual tuning for each new dataset, nnU-Net automatically adapts its architecture, preprocessing steps, training parameters, and post-processing strategies based on the characteristics of the input data [11].

RA-UNet combines residual connections with attention mechanisms in a 3D U-Net architecture, improving segmentation accuracy for liver and brain tumors. The residual attention modules help the network focus on relevant features while maintaining the benefits of deep architectures [12].

LATUP-Net is a lightweight 3D Attention U-Net with parallel convolutions somewhat similar to the inception model. It is designed for efficient brain tumor segmentation. It incorporates parallel convolutions to capture multi-scale information and attention mechanisms to refine feature representations, achieving high performance with reduced computational requirements [13].

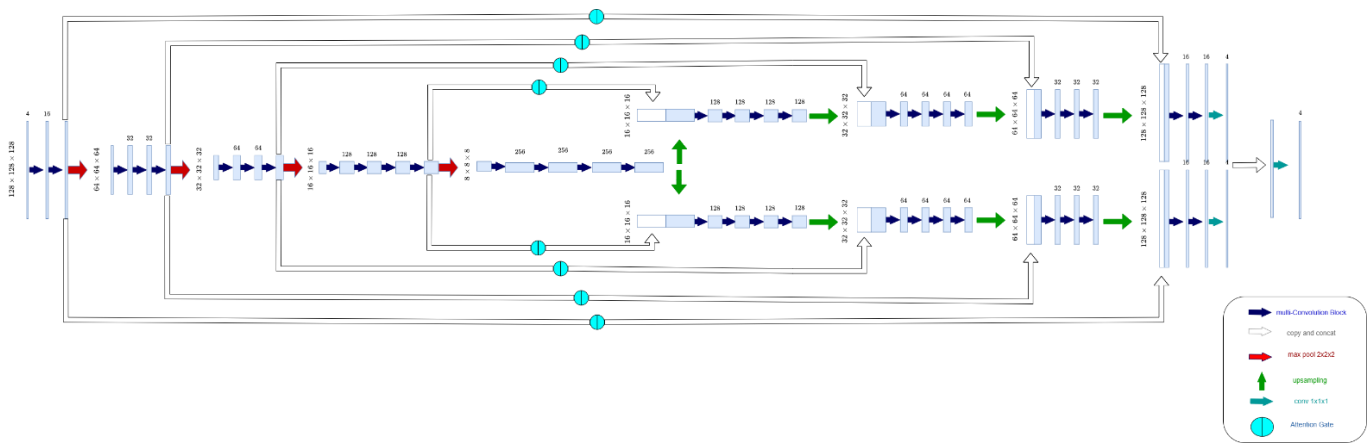

*Figure 3. Architecture of the proposed Dual-Decoder Attention U-Net. The model consists of a shared encoder followed by two parallel decoder branches. Each decoder processes feature maps using its own attention gates to selectively focus on salient regions from the encoder. Outputs from both decoders are combined to produce the final segmentation. The encoder employs a series of convolutional blocks with downsampling via max pooling, while each decoder performs upsampling and concatenation with attention-modulated skip connections. This design enables specialized decoding and improved feature refinement for brain tumor segmentation.*

## 3. Methodology

### 3.1 Overview

Following extensive experimentation with different architectures, strategies, and hyperparameters, we propose a customized U-Net variant for brain tumor segmentation highlighted in figure 3. The final model incorporates two parallel decoders instead of one, with each decoder connected to the encoder via its own set of attention gate mechanisms. These gates selectively pass spatial features deemed relevant to the decoder, suppressing less informative regions. Unlike the original Attention U-Net where decoder features from lower levels are used to guide higher-level encoder features, our design connects encoder and decoder features at the same spatial level. This modification aims to preserve spatial alignment while enabling each decoder to learn potentially distinct representations due to the stochastic nature of training. The outputs of both decoders are concatenated and passed through a final convolutional layer to produce the final segmentation map.

The model is trained using a combination of Dice loss and Focal loss, with a dynamic learning rate schedule. To improve generalization, we apply 3D Dropout as regularization and use weight decay. We utilize the BraTS 2020 dataset, which includes four input MRI modalities (T1, T1CE, T2, and FLAIR), and produce multi-class segmentation maps with five label values (0–4), one of which is missing (label 3).

### 3.2 Preprocessing

The BraTS 2020 dataset provides four MRI modalities per subject—T1, T1CE, T2, and FLAIR—and includes corresponding segmentation masks with five labels (0 to 4), where label 3 is unused and thus removed from training. The images are stored in volumetric format, which are first read and converted to NumPy arrays. All four modalities are stacked along the channel dimension to form a single input tensor.

Since the original images are large ($240 \times 240 \times 155$), a central cropping strategy is applied to remove background regions and reduce computational load. Unlike resizing, cropping better preserves tumor shapes and spatial resolution. Finally, the cropped volume size is $128 \times 128 \times 128$.

Segmentation masks are one-hot encoded to represent four effective classes. The total dataset contains 369 subjects, split into 75% training (257 subjects) and 25% testing (86 subjects).

Each input image undergoes z-score normalization, applied per sample, using its own mean and standard deviation. To improve generalization, several random augmentations are applied to each training sample with a probability of 0.2:

- Horizontal flipping
- Rotation (−10° to 10°)
- Intensity scaling (0.9–1.1)
- Gaussian noise ($\sigma = 0.01$)

The preprocessing pipeline is implemented using PyTorch's Dataset and DataLoader classes, which allows efficient batching, shuffling, and on-the-fly transformation of data during training.

### 3.3 Model Architecture

We propose a 3D U-Net variant with dual decoders and attention gates for brain tumor segmentation. The architecture is designed to promote diverse feature learning across independent decoder paths, while spatially refining encoder-decoder skip connections through dedicated attention mechanisms.

**Encoder**

The encoder path follows a typical U-Net structure with five hierarchical stages. Each stage includes a multi-convolution block comprising 2 to 4 3D convolutional layers, depending on the depth. Each convolution is followed by Group Normalization and ReLU activation. The choice of Group Normalization [14], which divides channels into groups and normalizes each group based on its statistics instead of Batch normalization [15] which normalizes a batch using the batch statistics, was made because of small batch size of 1 to promote stability. The feature dimensionality increases progressively:

$$\text{Input}=4 \rightarrow 16 \rightarrow 32 \rightarrow 64 \rightarrow 128 \rightarrow 256.$$

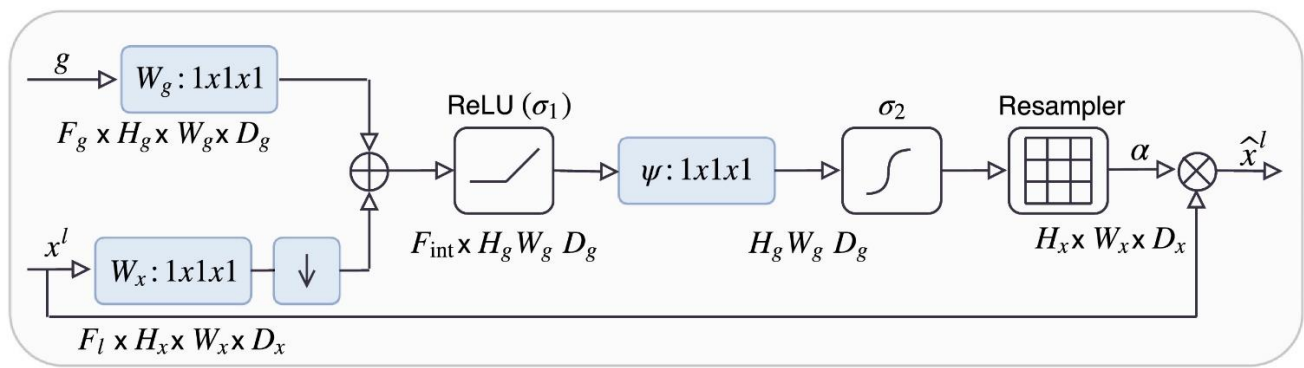

*Figure 4 Original Attention Gate (AG) mechanism proposed by Oktay et al. (Attention U-Net) [9]. The attention gate takes two inputs: a gating signal $g$ from a deeper decoder layer and a skip connection $x^l$ from the encoder. Each input is passed through a $1 \times 1 \times 1$ convolution to reduce dimensionality. The resulting feature maps are summed and passed through a ReLU activation followed by another $1 \times 1 \times 1$ convolution and a sigmoid activation to produce the attention coefficients $\alpha$. These coefficients are used to weight the encoder features.*

*In the original architecture, a resampling operation is applied to align the spatial dimensions of the gating and encoder signals. However, in our proposed model, the gating signal and encoder features originate from the same resolution level of the U-Net, making the resampling step unnecessary.*

**Attention Gates**

To enhance skip connections, attention gates are used independently for each decoder branch. Each gate computes attention coefficients for the corresponding encoder features, conditioned on decoder activations. The formulation is based on additive attention.

In the original Attention U-Net design (Oktay et al., 2018) [9], which is shown in figure 4, attention gates are employed at the skip connections between the encoder and decoder. Notably, the gating signal is derived from a deeper, lower-resolution decoder feature map, which is used to modulate the corresponding encoder features before merging. This approach is effective in suppressing irrelevant background information but may reduce the fine-grained spatial detail available at the corresponding encoder level.

In our proposed model, the attention mechanism that operates at the same spatial resolution yielded better results. In other words, the gating signal is computed directly from the decoder features that correspond in resolution to the encoder features. Both inputs—namely, the encoder skip connection (X) and the decoder signal (G)—are projected via $1\times1\times1$ convolutions into a common feature space. Their sum passes through a ReLU and then a $1\times1\times1$ convolution with a sigmoid activation to generate an attention map. This map is used to reweight the encoder features before they are concatenated with the decoder output. By using same-level gating, our design seeks to retain finer spatial context and thereby capture more localized features, potentially enhancing the segmentation of intricate tumor boundaries.

**Dual Decoders**

The network includes two independent decoder branches:

- Each decoder upsamples features using ConvTranspose3D.
- The upsampled features are concatenated with the attention-weighted encoder features.
- A multi-convolution block (same format as encoder) is applied to refine the merged features.

Each decoder outputs a segmentation map through a final $1\times1\times1$ convolution.

**Fusion and Output**

The outputs from the two decoders are concatenated along the channel axis and passed through a final 1×1×1 convolution layer to produce the final segmentation logits. This fusion step enables the model to leverage complementary features extracted by each decoder.

**Normalization and Regularization**

All convolution blocks use Group Normalization [14] (due to small batch size = 1) which performs better on small batch sizes.

Dropout3D is used after each convolution block:

- Dropout rate = 0.1 for channels ≤ 32
- Dropout rate = 0.2 for channels ≤ 128
- Dropout rate = 0.3 for channels > 128

Dropout3D drops whole features instead of dropping individual neurons. This keeps the spatial integrity of features intact which is useful for the case of tumors with small sizes.

**Loss Function**

To effectively address class imbalance and promote accurate delineation across tumor subregions, we employ a composite loss function combining multi-class Dice Loss and multi-class Focal Loss. This hybrid formulation allows the model to capture both global overlap and focus on hard-to-classify voxels for each tumor class.

1. Multi-Class Dice Loss

Dice Loss is an overlap-based metric that measures similarity between the predicted segmentation and ground truth. For multi-class segmentation, we compute Dice Loss independently for each class and then average the results. This loss ensures balanced training across all classes, including small tumor subregions. It is formulated as:

$$L_{Dice} = 1 - 2 \times \frac{P \cap T}{P + T}$$

2. Multi-Class Focal Loss

Focal loss is a modified version of cross-entropy loss that weights hard and easy samples (voxels) differently. Easy voxels are those that are classified correctly with a high probability and hard voxels are those that are misclassified with higher probability.

Focal loss reduces the impact of well-classified voxels and amplifies the gradient from hard-to-classify regions. It thereby encourages the model to concentrate on underrepresented or ambiguous regions that are often misclassified. It is formulated as:

$$L_{Focal} = -\sum_{n=1}^{N} -\alpha(1-p_t)^{\gamma}\log(p_t)$$

In which $p_t = p$ if the ground truth $y = 1$ and $p_t = 1 - p$ if $y = 0$. $\gamma$ is the focusing parameter and is set to 2 in our implementation. $\alpha = 0.25$ is the class-balancing parameter.

Final Loss

We combine the two loss terms as:

$$L_{total} = \lambda_1 \times L_{Dice} + \lambda_2 \times L_{Focal}$$

In our setup, we use $\lambda_1 = 0.7, \lambda_2 = 0.3$

### 3.4 Training Procedure

We trained our model using the AdamW optimizer with AMSGrad and weight decay to promote generalization. A OneCycle learning rate policy with cosine annealing was employed to improve convergence. The loss function combined Dice Loss and Focal Loss to address class imbalance in the BraTS dataset.

During training, we computed per-class Dice scores, sensitivity, specificity, and region-based metrics for Whole Tumor (WT), Tumor Core (TC), and Enhancing Tumor (ET) to monitor segmentation quality.

After each epoch, validation was conducted to assess generalization. The model was evaluated using the same metrics as in training, including Dice scores and region-wise segmentation performance for WT, TC, and ET. No gradients were computed during validation.

Training and validation were carried out using a GTX1080 GPU with 8GB vram.

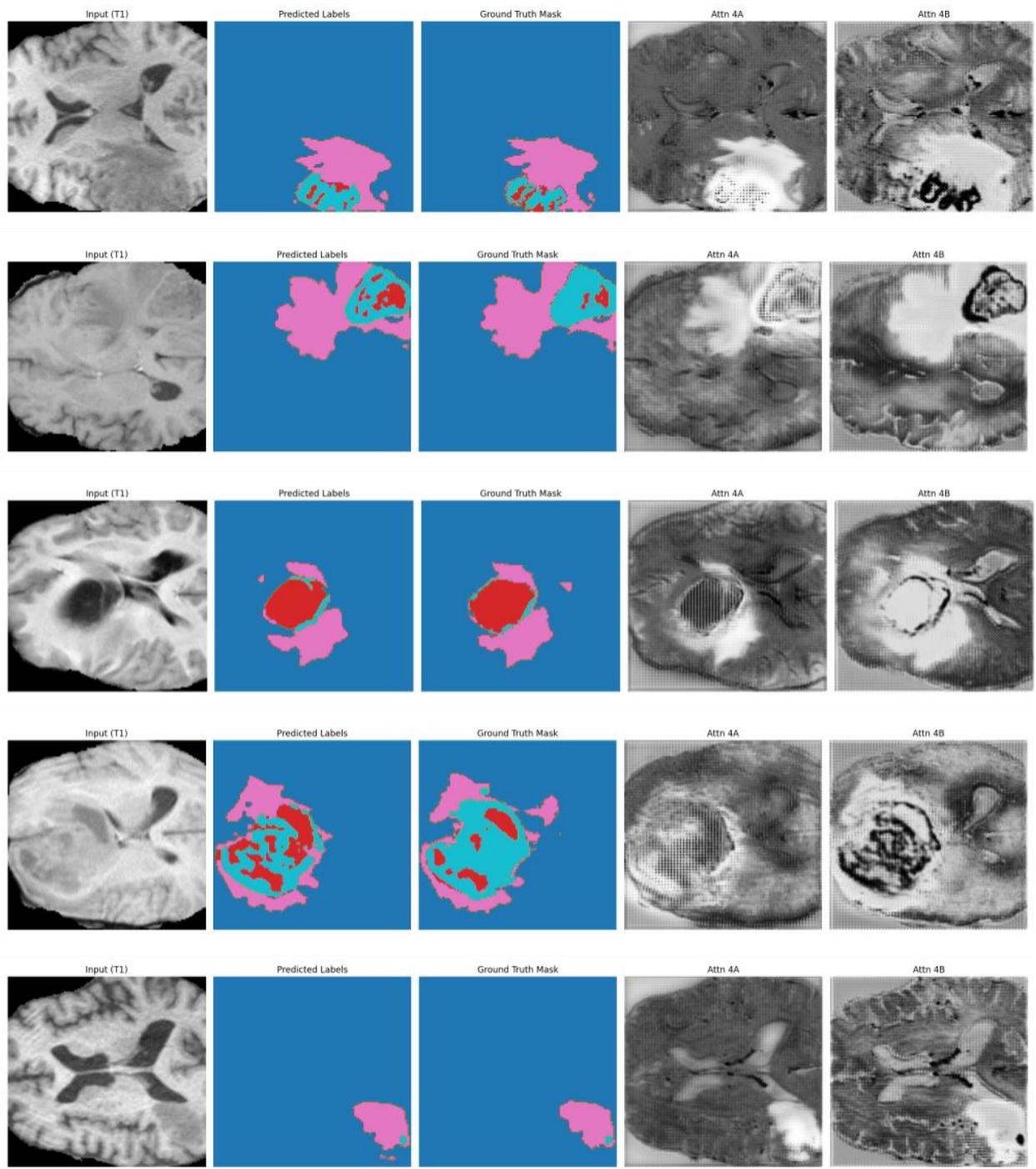

*Figure 5 Results of the proposed dual-attention-gated U-Net model on validation samples from the BraTS 2020 dataset. Each row corresponds to one sample and shows (from left to right): the T1-weighted input image, the predicted segmentation labels, the ground truth mask, and the output attention maps from the two attention gates in the final decoder layer (Attn 4A and Attn 4B).. The attention maps demonstrate how the dual attention gates learn complementary spatial features: Attn 4A often highlights broader contextual areas and background suppression, while Attn 4B tends to focus more sharply on localized tumor boundaries and enhancing regions. This complementary behavior enables the model to more effectively capture both global structure and fine-grained tumor details.*

## 4. Experiments and Results

### 4.1 Comparison between architectures

Among the architectures tested, the most prominent ones are presented in Table 1. The baseline UNet represents the original architecture. The 3AG+TDUNet model incorporates three separate attention gates with three decoders, to test whether increasing decoder branches and attention mechanisms improves performance. The 1AG+ResDDUNet model adds residual connections to the convolutional blocks of a dual-decoder network with a single attention gate. Another novel architecture, the 1AG+biDDUNet, introduces a bidirectional flow of information. In this model, encoder features are influenced by decoder features from the previous iteration. This is achieved by saving decoder layer outputs at each level and feeding them back to the encoder at the same level in the next iteration through concatenation or addition.

The final and proposed model is the 2AG+DDUNet, which includes dual decoders and two attention gates. This model outperformed the other architectures across most metrics as shown in figure 6. Since the focus of this work is to balance efficiency and accuracy, more complex structures that did not yield performance improvements are omitted. For instance, models incorporating self-attention blocks—with or without attention gates—after convolutional layers did not improve performance and significantly increased computational costs.

The results of running the model on validation data is demonstrated in figure 5 for qualitative analysis.

The models were evaluated based on three main metrics shown below:

$$Dice = 2 \times \frac{P \cap T}{P + T}$$

In which $T$ is our ground truth and $P$ is the predicted mask.

$$Sensitivity = \frac{TP}{TP + FN}$$

$$Specificity = \frac{TN}{TN + FP}$$

In which $TP$ is the true positive, $TN$ is the true negative, $FP$ is false positive and $FN$ is false negative.

Table 1 Performance comparison of U-Net variants on the BraTS 2020 dataset.

(Evaluated using Dice score, sensitivity, and specificity for WT, TC, and ET.)

| Model name | Dice (WT) | Dice (TC) | Dice (ET) | Sensitivity (WT) | Sensitivity (TC) | Sensitivity (ET) | Specificity (WT) | Specificity (TC) | Specificity (ET) |
|---|---|---|---|---|---|---|---|---|---|
| UNet | 84.75 | 78.54 | 66.63 | 85.53 | 80.24 | 73.96 | 99.35 | 99.75 | 99.77 |
| 3AG+TDUNet | 83.83 | 77.60 | 67.59 | 87.65 | 75.71 | 69.02 | 99.10 | 99.76 | 99.85 |
| 1AG+ResDDUNet | 81.58 | 72.63 | 62.01 | 82.13 | 77.73 | 69.29 | 99.29 | 99.61 | 99.70 |
| 1AG+biDDUNet | 80.13 | 72.66 | 62.84 | 83.29 | 77.89 | 71.19 | 98.99 | 99.50 | 99.65 |
| 2AG+DDUNet | 85.06 | 80.61 | 71.26 | 84.85 | 79.64 | 71.55 | 99.46 | 99.76 | 99.83 |

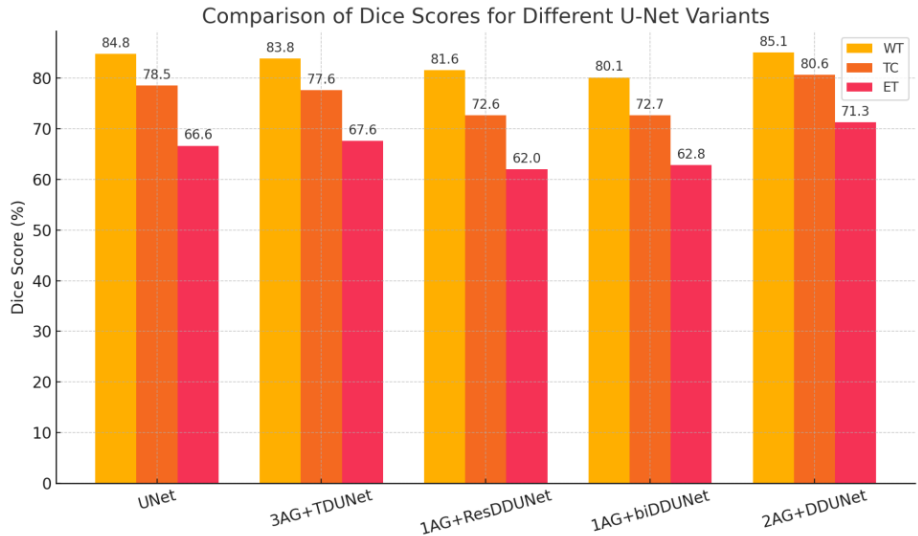

Figure 6 Comparison of Dice scores for Whole Tumor (WT), Tumor Core (TC), and Enhancing Tumor (ET) across different U-Net-based architectures evaluated on the BraTS 2020 dataset. The proposed dual-decoder U-Net with two attention gates (2AG+DDUNet) achieved the highest performance in all three categories.

### 4.2 Ablation study

An extensive ablation study was conducted to analyze the impact of various architectural and training modifications. Key results are summarized in Table 2 and visualized in figure 7.

After identifying the 2AG+DDUNet as the best-performing model, further experiments were conducted to optimize training strategies. One key finding was that a dynamic learning rate significantly outperformed a constant one. Two learning rate schedulers were compared: OneCycle (OC_lr) and Cosine Annealing with Warm Restarts (CAWR_lr). The OneCycle schedule consistently yielded better performance, and was therefore used in subsequent experiments.

Several architectural variations were also tested:

- 1AG+DDUNet+OC_lr: A model with a shared attention gate and the OneCycle scheduler.
- 1AG+DDUNet+CAWR_lr: The same model using the CosineAnnealingWarmRestart scheduler.
- 2AG+DDUNet (dist_map): A variant where one decoder predicts a distance map to improve boundary detection. The decoder was trained using MSE loss between the predicted and ground truth distance maps.
- 1AG (org)+DDUNet: Implements the original attention gate mechanism, which uses deeper decoder features, instead of same-level ones as in the proposed method.
- 2AG+DDUNet+SC: A version that uses strided convolution for upsampling to test if learnable upsampling improves over standard max pooling.
- 2AG+SE+DDUNet: Adds Squeeze-and-Excitation (SE) blocks to assess the effect of channel-wise attention.

*Table 2 Ablation study results for architectural and training modifications.*

*(Includes the impact of learning rate schedules, attention mechanisms, and upsampling strategies.)*

| Model name | Dice (WT) | Dice (TC) | Dice (ET) | Sensitivity (WT) | Sensitivity (TC) | Sensitivity (ET) | Specificity (WT) | Specificity (TC) | Specificity (ET) |
|---|---|---|---|---|---|---|---|---|---|
| 1AG+ DDUnet +OC_lr | 84.82 | 80.26 | 67.71 | 84.55 | 78.39 | 66.16 | 99.43 | 99.77 | 99.89 |
| 1AG+ DDUnet +CAWR_lr | 83.10 | 75.07 | 66.22 | 82.04 | 71.21 | 67.01 | 99.41 | 99.77 | 99.82 |
| 2AG+ DDUNet (dist_map) | 84.54 | 78.33 | 68.45 | 85.99 | 75.48 | 69.51 | 99.28 | 99.82 | 99.87 |
| 1AG (org)+ DDUnet | 82.47 | 79.91 | 66.23 | 84.30 | 73.93 | 67.92 | 99.21 | 99.73 | 99.81 |
| 2AG+ DDUNet+ SC | 84.83 | 79.52 | 67.73 | 87.89 | 80.21 | 71.89 | 99.25 | 99.74 | 99.82 |
| 2AG+SE+ DDUNet | 84.42 | 79.40 | 67.28 | 86.42 | 76.80 | 65.74 | 99.24 | 99.77 | 99.87 |
| 2AG+ DDUNet | 85.06 | 80.61 | 71.26 | 84.85 | 79.64 | 71.55 | 99.46 | 99.76 | 99.83 |

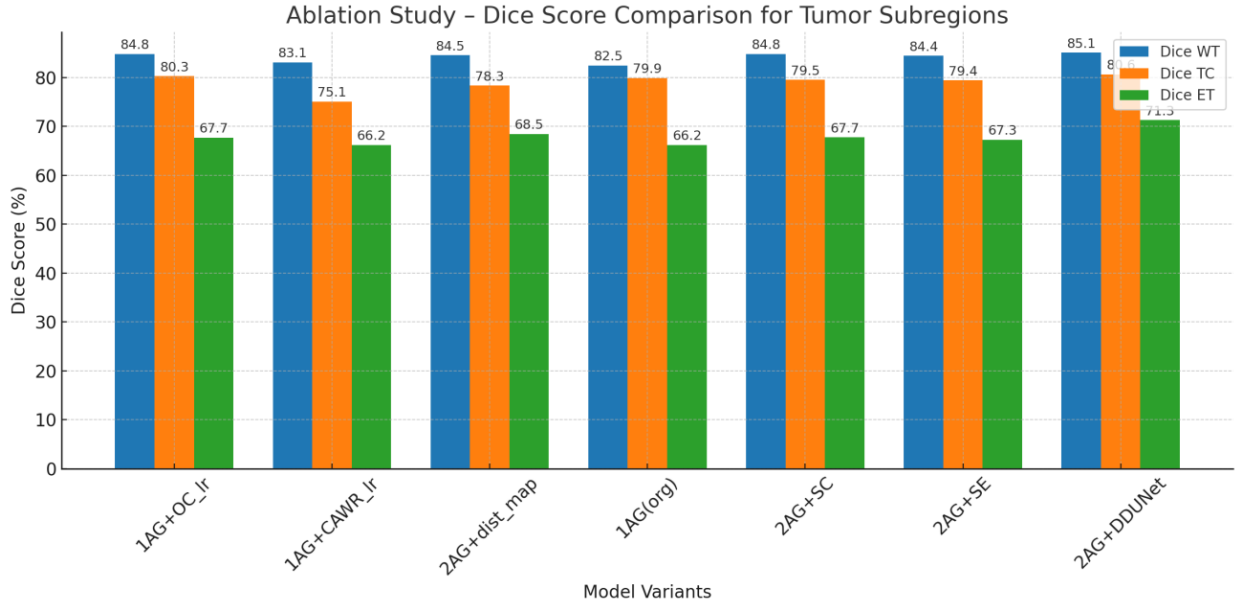

*Figure 7 Ablation study comparing the Dice scores of different architectural and training modifications to the dual-decoder U-Net. The 2AG+DDUNet configuration consistently outperformed other variants.*

## 5. Conclusion

This work introduced an attention-guided dual-decoder 3D U-Net architecture for brain tumor segmentation on the BraTS 2020 dataset. The proposed model processes four MRI modalities (T1, T1ce, T2, and FLAIR) as input channels and utilizes an encoder to extract hierarchical features while reducing spatial resolution for computational efficiency. Two separate decoders then reconstruct the segmentation map, each equipped with its own attention mechanism to selectively integrate encoder features.

This dual-path decoding strategy enables the model to learn diverse and complementary representations, improving segmentation accuracy across tumor subregions. Extensive experimentation with over 100 architectural variations and training strategies led to the selection of the optimal configuration. Despite being lightweight and trained for only 50 epochs on modest hardware, the model achieves state-of-the-art performance, outperforming conventional and attention-augmented U-Net variants in terms of Dice score, sensitivity, and specificity.

In summary, the proposed model balances performance and efficiency, achieving Dice scores of 85.06, 80.61, and 71.26 for Whole Tumor, Tumor Core, and Enhancing Tumor respectively—surpassing other variants while maintaining a small computational footprint. These results demonstrate the model's potential utility in clinical decision-support systems, particularly in low-resource environments.

## 6. Future Directions

This work introduces an efficient yet robust architecture for the task of brain tumor segmentation. The results could improve either by finding even more efficient and stronger architectures, leveraging better hardware allowing longer and more rigorous training setups.

While this study focuses on architectural optimization under constrained resources, future research could benefit from exploring techniques beyond supervised learning. One promising direction is the incorporation of self-supervised or semi-supervised learning to leverage large-scale unlabeled data. These methods can enhance model generalization, particularly for tumor regions with highly variable shapes, sizes, and locations.

## 7. Limitations

Although the proposed model achieves strong performance, several limitations remain.

First, segmentation of brain tumors is inherently complex due to the heterogeneous and irregular nature of tumor shapes and boundaries. Through our experiments, we observed a performance ceiling that cannot be overcome by architectural improvements alone. This "glass wall" likely stems from the limited size and diversity of labeled training data, which restricts the model's ability to generalize to difficult or rare tumor cases.

Second, the model was trained and evaluated solely on the BraTS 2020 dataset. While this is a widely used benchmark, it does not capture the full spectrum of variations found in real-world clinical imaging (e.g., different MRI protocols, noise levels, or institutions). Generalization to external datasets remains untested.

Furthermore, although efficient, the model is still not fully autonomous and should not be used without human oversight, particularly in high-risk clinical scenarios. It is best viewed as an assistive tool to enhance human decision-making, especially in settings where expert availability is limited.

Lastly, hardware constraints limited the scope of training and experimentation. Despite this, the results are encouraging and suggest that performance can be further improved with more computational resources and extended training schedules.

**Acknowledgment**

This research was conducted independently. The author would like to thank Hossein Danesh Pajouh for his encouragement and providing the computer used in this research.

During the preparation of this work the author used ChatGPT in order to improve language and readability. After using this tool, the author reviewed and edited the content as needed and takes full responsibility for the content of the publication.